\documentclass[a4paper,10pt]{article}
\usepackage[utf8]{inputenc}
\usepackage{url}
\usepackage{mathptmx,mathtools}
\usepackage{anyfontsize}
\usepackage{t1enc}
\usepackage{amsmath,amssymb,fourier-orns}
\usepackage{blindtext}
\usepackage{multicol,changepage}
\usepackage{color}
\usepackage{cancel}
\usepackage{relsize}%
\usepackage{mathrsfs}
\usepackage{graphicx}
\usepackage{framed}
\usepackage[margin=1cm]{caption}
\usepackage[english]{babel}
\usepackage{amsthm}

\providecommand{\keywords}[1]
{
  \small	
  \textbf{\textit{Keywords---}} #1
}

\DeclarePairedDelimiterX{\expectarg}[1]{[}{]}{%
  \ifnum\currentgrouptype=16 \else\begingroup\fi
  \activatebar#1
  \ifnum\currentgrouptype=16 \else\endgroup\fi
}

\newcommand{\innermid}{\nonscript\;\delimsize\vert\nonscript\;}
\newcommand{\activatebar}{%
  \begingroup\lccode`\~=`\|
  \lowercase{\endgroup\let~}\innermid 
  \mathcode`|=\string"8000
}

\newtheorem{theorem}{Theorem}
\title{A Novel Solution for the General Diffusion}
\author{Luisiana Cundin}
\begin{document}

\maketitle

\begin{abstract}
The Fisher\textendash KPP equation is a reaction\textendash diffusion equation originally proposed by Fisher to represent allele propagation in genetic hosts or population. It was also proposed by Kolmogorov for more general applications. A novel method for solving nonlinear partial differential equations is applied to produce a unique, approximate solution for the Fisher\textendash KPP equation. Analysis proves the solution is counter\textendash intuitive. Although still satisfying the maximum principle, time dependence collapses for all time greater than zero, therefore, the solution is highly irregular and not smooth, invalidating the traveling wave approximation so often employed. 
\end{abstract}

\keywords{KPP equation or Fisher–KPP equation}



\section{On the general diffusion\ldots}

Diffusion is incontrovertibly one of the most fundamental physical processes describing the transport of energy, mass and other phenomena through space and time. The linear transfer of energy is certainly in the purview of Nature's design, as evidenced all around us, everyday; but, nonlinear transfer has long been sought as an extension to rational systems where the medium reacts to the gradient of concentration. A nonlinear diffusive process would describe some process that, in addition to gradient dependence, would be further dependent upon the medium response to the gradient of concentration, reacting in some nonlinear, chaotic fashion \cite{Fisher,KPP1937}. The concept of some unusual medium response is predicated upon the idea that the medium carries some special reactive properties, like memory or retention or something of the sort that would account for some additional behavior atop ordinary flux allowed by the medium. 

It may be that any such nonlinear diffusive behavior could simply be captured in a modified diffusion constant, rather, experimentalists may find unique diffusion constants for various media, and, never think it due to some reactive medium response \cite{Ahuja,Hou,Cherniha_2015}. In fact, any exotic, complex diffusivity should be focused upon the diffusion constant, \emph{per se}, and not extricated to describe the medium response itself. In general, the media response is typically random in nature, as \emph{per} Brownian motion; moreover, although randomness is often perceived nonlinear, it is rational nonetheless \cite{Miyaguchi}. Whether it is physically possible for media to exhibit nonlinear reactive responses is certainly questionable, in fact, it may all together be not possible for such systems to exists; because, linear diffusion already describes the flux of energy and mass subject to the local gradient or concentration, therefore, already accounts for the actual medium response in its description of mass/energy transfer \cite{Lenzi}. To ascribe additional ''nonlinear'' attributes to media would imbue into physical material\'{e} some ghost process or special physical phenomenon yet discovered. 

In many cases, for example,  reaction\textendash diffusion processes, as are certain chemical reactions, nonlinear diffusion equations are used in an attempt to capture the media's actual response to changes in the gradient as the chemical reaction progresses in time and space, therefore, there is some rational\'{e} behind attempting to formulate some type of equation that mimics a very complex, intricate play in reversible concentrations, as the chemical reaction progresses to homeostasis \cite{Britton}. Nevertheless, to formulate or reduce such a complex process to one, all encompassing equation may or may not be suitable, adequate or advisable, for the process is truly an interplay of many reaction chemical kinetic equations, interacting over time and space, with an intricate dance in concentrations diminishing and evolving in time. Complex systems, generally speaking, require a milieu of equations, each formulated specifically, to properly model each element/species involved in the overall process \cite{Stundzia}. In this manner, a researcher adequately accounts for each kinetic species in the reaction and does not attempt to 'cut corners' by reducing the entire system to some modified ''nonlinear'' system, which is often forced, in some manner, into a one\textendash dimensional system; this done in hopes it will mimic the system in question and admit ease in solving. 

There are many reasons the simple addition of some multiplicative medium response is not advisable nor should be considered physically correct. Of the many reasons, Ostrogradsky's theorem concerning mechanics stands tall \cite{Ostrogradsky}. The theorem shows why no system with higher order derivatives than second order are physically acceptable or seen in mechanics, furthermore, it proves third order and higher systems are unstable in time and space, therefore, describe unrealistic physical processes, or, in the least, doomed processes. All systems are bounded from above by energy and by material properties (strengths); but, to say a system's Hamiltonian is unbounded from below envisions a system which allows a particle at rest to suddenly jump from rest to some unbounded state of energy, moreover, a negative state of energy, which is improper, to say the least. All energy is positive, although, in a relative system or coordinate system, negative energies are possible, even though, such formulations are sloppy, at best, but improper, at worst. 

Nonlinear equations admit special solutions, usually, involving hyperbolic trigonometric equations, which are typically periodic functions and suffer from branch points, singularities and other non-analytic properties. The fact poles exist within the domain of the solution should indicate to the wary that the domain is not entire, therefore, the solution is hemmed in or restricted in some way or another. A typical path for solving nonlinear equations involves the traveling wave approximation, which assumes any deflection in the solution is small enough and smooth enough to allow for equivalence under  D'Alembert's principle, thereby, enabling reduction of an equation of some dimension to a one\textendash dimensional canonical equation, whose behavior is assumed equivalent \cite{Glocker,Coopersmith}. Nonlinear equations, in general, are not smooth and this fact brings any solution obtained via the traveling wave approximation under serious scrutiny, as will be seen for the Fisher\textendash KPP equation, shortly. The additional multiplicative medium response places extraordinary restraints upon the system and any solution. 

\section{Fisher-KPP equation}

Consider the general Fisher\textendash KPP equation with real coefficients (\textit{D, b, r}), where \textit{D} describes the diffusivity of the material under question, \textit{b} the linear response of the medium (if applicable), and \textit{r} the nonlinear coefficient, \emph{viz.:}

\begin{equation}\label{main}
 \frac{\partial}{\partial t}u(x,t)=D \frac{\partial^2}{\partial x^2}u(x,t)-b\;u(x,t)+r\;u(x,t)^2/;\;\{(D,b,r)|(D,b,r)\in\Re\}
\end{equation}

\vspace{5pt}

Assume a solution in the form of $u(x,t)=G(x,t)\ast f(x,t)$, where $G(x,t)$ represents Green's function which solves the linear equation, \textit{inept}, $r=0$; \emph{viz.:} 
\begin{equation}
 G(x,t)=\frac{e^{-x^2/4Dt}}{\sqrt{4\pi D t}}\subset e^{-(2\pi s)^2 D t}e^{-bt}=g(s,t),
\end{equation}
\noindent where symbol $\subset$ is borrowed from Bracewell, indicating the Fourier transform of the given function \cite{Bracewell}. 

Since Green's function is a solution to the linear partial differential equation with linear medium response already included, then the task at hand is to reduce the main equation to some residual equation involving only the nonlinearity. Employing Green's solution realizes several cancellations immediately, cancelling several terms after applying the time derivative, where partial derivatives distribute across convolution integrals, see Theorem \ref{theorem2}. Moreover, one retains the choice onto which function a derivative is acted upon, if the convolution integral involves the same variable, to wit, the choice has been made to apply both spatial derivatives to Green's function, \emph{viz.:}

\begin{equation}\label{intermediate}
 \cancel{(G'\ast f)(x)}+(G\ast f')(x)=\cancel{D\,\left(G_{xx}\ast f\right)(x)}-\cancel{b\;\left(G\ast f\right)(x)}+r\;\left(G\ast f\right)^2(x),
\end{equation}
\noindent where prime indicates derivative with respect to time. 

Since all terms are removed except the nonlinear term and the corresponding time derivative needed to cancel this term, the residual equation, \emph{viz.:}
\begin{equation}\label{solve}
 G\ast f'=r\;\left(G\ast f\right)^2(x)
\end{equation}

No immediate solution comes to mind for the above equation, so, a sequence of approximations for the convolution integral will be employed, namely,
\begin{theorem}[Convolution integral theorem]\label{conv1} The convolution integral is greater than or equal to the product of both functions involved in the convolution integral
 \begin{equation}\label{approx}
 f\ast g\ge fg
\end{equation}
\end{theorem}

Theorem \ref{conv1} is true in general because the convolution integral sums an area either larger than or equal to the area of both functions multiplied together. This approximation enables bypassing the convolution integral and solving the resulting ordinary differential equation.

\section{Zeroth approximation}
With the aid of the approximation for a convolution integral, theorem \ref{conv1}, the residual nonlinear equation can be transformed into a Bernoulli equation, which has an exact solution. Applying the approximation once will yield interesting results and they will be investigated.

Firstly, equation (\ref{solve}) will be brought to bear in the frequency domain to yield a zeroth approximation. In the original domain, the approximation must be applied to two convolution integrals, whereby, in the codomain, the approximation is only for one convolution integral, therefore, the overall approximation should be closer. In the transform domain, the residual nonlinear equation takes on the following form, plus, the approximation, after applying theorem \ref{conv1}, is to the right of the inequality, \emph{viz.:}
\begin{equation}
 gF_t=r\left(gF\ast g F\right)(s)\ge gF_t=rg^2F^2
\end{equation}

The approximation initiates a set of applications to yield a solution, \emph{viz.:}
\begin{align*}
 gF_t=& rg^2F^2,\\
 \frac{F_t}{F^2}=& rg,\\
 -\frac{1}{F}= & C(s)+r\int{g\,\mathrm{d}t},\mathrm{ finally,}\\
 F=& \frac{1}{C(s)-r\int{g\,\mathrm{d}t}},
\end{align*}
\noindent where $C(s)$ is a constant of integration. 

The solution states the following, in the frequency domain, \emph{viz.:}
\begin{equation}
 u(s,t)=gF
\end{equation}

The solution, as it stands now, is quite acceptable, for it does reduce to the solution to the linear equation for nonlinear coefficient \textit{r} equal to zero; assuming the constant of integration, $C(s)$, equals unity. 

A general inverse Fourier transform to the original domain is prohibitive, given the nature of the solution, but an expansion around small nonlinear coefficient \textit{r} yields an approximate solution that satisfies what many researchers yearn for from the Fisher-KPP equation; since most applications of the Fisher-KPP equation involve small nonlinear coefficient \textit{r}. Since the denominator of the function \textit{F} involves the integral of the Green's function in the transform domain, this function's maximum is unity and declines as either the frequency increases or time increases, therefore, for small nonlinear coefficient \textit{r}, the requirements for a binomial expansion are satisfied.  

Before expanding, it is advisable to define what the constant of integration is, and, for the purposes of this monograph, the property that the solution approach the Dirac Delta function in space as time approaches zero ensures the solution is normalized on the domain. To that end, setting time equal to zero in the finction \textit{F} yields a constant of integration equal to $C(s)=1-r/((2\pi s)^2D+b)$, which will effectively cancel the contribution from the algebraic term in the limit of time to zero, yet enable a bounded formul\ae.

\begin{figure}[t]
\centering
 \includegraphics[scale=0.75]{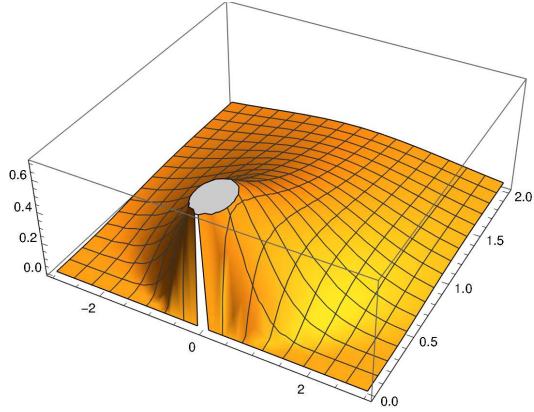}
 \caption{Zeroth approximation after binomial expansion (gray area is clipped by software \cite{mathematica}) for $D=1,b=1,r=0.1$ and domain $\mathcal{D}\subset\{x|x=(-3,3)\},\{t|t=(0,2)\}$.}
 \label{Fig1}
\end{figure}

Lumping together all functions under the symbol zeta, $\zeta$, the following relationship holds for the binomial approximation:
\begin{equation}
 F\approx\frac{1}{1-r\;\zeta}=\sum_{n=0}^\infty{(r\,\zeta)^n}
\end{equation}

The inverse Fourier transform of the first term of the expansion is shown for small nonlinear coefficient (\textit{r}):
\begin{equation}
 u(s,t)=gF\approx e^{-(2\pi s)^2Dt-bt}-\frac{r\,e^{-(2\pi s)^2Dt-bt}}{(2\pi s)^2D+b}+\frac{r\,e^{-2(2\pi s)^2Dt-2bt}}{(2\pi s)^2D+b}+\text{higher order terms}
\end{equation}

All relevant inverse Fourier transforms are shown below:
\begin{align}
e^{-(2\pi s)^2Dt-bt}&\supset \frac{e^{-\frac{x^2}{4Dt}}}{\sqrt{4\pi Dt}}e^{-bt}\\
 \frac{1}{(2\pi s)^2D+b} &\supset\frac{ e^{x \sqrt{\frac{b}{D}}}}{2 \sqrt{b D}}\theta (-x)+\frac{ e^{-x
   \sqrt{\frac{b}{D}}}}{2 \sqrt{b D}}\theta (x)\\
   \frac{e^{-(2\pi s)^2Dt}e^{-bt}}{(2\pi s)^2D+b}&\supset\frac{ e^{\frac{b x}{\sqrt{b D}}}e^{b t} \text{erfc}\left(\frac{2 t \sqrt{b
   D}+x}{2 \sqrt{D t}}\right)\theta (-x)}{4 \sqrt{b D}}+\frac{ e^{b t}e^{-\frac{b x}{\sqrt{b
   D}}} \text{erfc}\left(\frac{2 t \sqrt{b D}-x}{2 \sqrt{D t}}\right)\theta (x)}{4 \sqrt{b D}}\\
   \frac{e^{-2(2\pi s)^2Dt}e^{-2bt}}{(2\pi s)^2D+b}&\supset\frac{ e^{\frac{b x}{\sqrt{2b D}}}e^{-b t} \text{erfc}\left(\frac{2 t \sqrt{2b
   D}+x}{2 \sqrt{2D t}}\right)\theta (-x)}{4 \sqrt{2b D}}+\frac{ e^{-b t}e^{-\frac{b x}{\sqrt{2b
   D}}} \text{erfc}\left(\frac{2 t \sqrt{2b D}-x}{2 \sqrt{2D t}}\right)\theta (x)}{4 \sqrt{2b D}},
\end{align}
\noindent where $\theta(x)$ is Heaviside's step function.

The zeroth approximation, under the binomial expansion, mimics the linear solution, hence, the modification is slight, indeed. The surface shown in Figure \ref{Fig1} is slightly depressed from the linear surface, hence, the effect of the nonlinear coeiffient is to compress and spread the energy faster than under normal circumstances. The solution still approaches a Dirac Delta in space in the limit of time to zero, ensuring a solution satisfying all requirements for a Green's function, therefore, the solution shown is entire and analytic throughout the domain $\mathcal{D}$. In addition, the solution satisfies all releavant boundary conditions, specifically, the solution is zero along the outer boundary of the domain $\partial\mathcal{D}$. 

\section{Successive approximations}
A single application of the approximation for convolution integrals, theorem \ref{approx}, provided a meaningful solution to the nonlinear reaction\textendash diffusion equation, equation (\ref{main}); nonetheless, it is only one application of the approximation and successive applications will reveal a more accurate solution. 

To that end, assume the following solution for equation (\ref{main}):
\begin{equation}
 u(x,t)=G\ast f_1\ast f_2
\end{equation}

Applying the time derivative and maintaining all spacial derivatives upon Green's function yields the following residual function in the frequency domain:
\begin{equation}
 gf_1'f_2+gf_1f_2'=r\left(gf_1f_2\ast gf_1f_2\right)(s)\ge gf_1'f_2+gf_1f_2'=rg^2f_1^2f_2^2,
\end{equation}
\noindent where, similarly, the approximation has been applied to the right of the inequality, also, prime represents derivation with respect to time. 

The prime of $f_1$ is known, it is simply the zeroth order approximation, after applying the derivative, $f_1'=-rgf_1^2$, which will aid in resolving the subsequent Bernoulli ordinary differential equation, where function $f_2$ is the unknown functional \cite{boyce1997elementary}, \emph{viz.:}

\begin{center}
 \begin{minipage}{0.75\textwidth}
 \begin{framed}
 \begin{center}
  Another application of the approximation
 \end{center}
 \begin{equation}
  gf_1'f_2+gf_1f_2'=rg^2f_1^2f_2^2
 \end{equation}

 Evaluating the prime against function $f_1$ yields:
 \begin{equation}
  -rgf_1^2f_2+f_1f_2'=rgf_1^2f_2^2,
 \end{equation}
 
 Then substituting $f_2=h^m$ and $f_2'=mh^{m-1}h'$, yields:
 \begin{equation}
  -rgf_1h^m+mh^{m-1}h'=rgf_1f_2^{2m}
 \end{equation}

With \textit{m}, Bernoulli's free parameter, equal to minus unity, the equation resuces to the following:
\begin{equation}
 rgf_1h+h'=-rgf_1
\end{equation}

Solving for the integrating factor:
\begin{equation}
 he^{\int{rgf_1\mathrm{d}\,t}}
 \end{equation}
 
which yields the following as a solution for the unknown function:
\begin{equation}
 f_2=\frac{e^{\int{rgf_1\mathrm{d}\,t}}}{C_2(s)-\int{rgf_1e^{\int{rgf_1\mathrm{d}\,t}}\mathrm{d}\,t}}
\end{equation}

 \end{framed}
\end{minipage}
\end{center}

Repeated iterations, as above, reveal the following sequence of functions:

\begin{align}
\begin{split}
f_1&=\frac{1}{C_1(s)-\int{r\,g\,\mathrm{d}\,t}}\\
 f_2& = \frac{e^{\int{r\,gf_1\,\mathrm{d}\,t}}}{C_2(s)-\int{r\,gf_1\,e^{\int{r\,gf_1\,\mathrm{d}\,t}}\mathrm{d}\,t}}\\
 f_3& = \frac{e^{\int{r\,gf_1f_2\,\mathrm{d}\,t}}}{C_3(s)-\int{r\,gf_1f_2\,e^{\int{r\,gf_1f_2\,\mathrm{d}\,t}}\mathrm{d}\,t}}\\
 &\hspace{50pt}\vdots\\
 f_{n+1} & =\frac{e^{\int{r\,gf_1f_2\ldots f_n\mathrm{d}\,t}}}{C_n(s)-\int{r\,gf_1f_2\ldots f_n\,e^{\int{r\,gf_1f_2\ldots f_n\,\mathrm{d}\,t}}\mathrm{d}\,t}}
 \end{split}
\end{align}

Inductive reasoning reveals the $f_{n+1}$ functional, whose limit to infinity approaches a functional $f_s$, \emph{viz.:}
\begin{equation}
\lim\limits_{n\rightarrow\infty}f_{n+1}\longrightarrow f_s/;\{f_s\subset\mathcal{D}\} 
\end{equation}

The functional $f_s$ is a product functional and has the property of equaling zero for all values of time and frequency variable greater than zero, otherwise, the functional equals unity in the limit of time to zero, hence, the functional behaves as a Dirac Delta functional in time, \emph{viz.:}
\begin{equation}
 f_s=\begin{cases} 
      \delta(t) & t>0 \\
      \delta'(t)=r\delta(t) & t=0
   \end{cases}
\end{equation}

\newpage

\section{Concluding remarks}
Much has been written on the subject of Fisher's equation in the literature, primarily, traveling wave solutions; albeit, such solutions, whether analytic or numeric solutions, if based upon the traveling wave approximation, are completely invalid. This conclusion was suspected by this author, given the Ostrogradsky theorem concerning mechanics.

The analysis shows the solution is not smooth and irregular, therefore, the traveling wave approximation should not be applied. Besides the serious implications for this particular equation studied, the Fisher-KPP equation, there are a host of nonlinear equation where the traveling wave approximation has been historically employed, placing solutions on shaky ground, to say the least. In fact, after analysis, a whole class of apparent published solutions have been placed under disbelief. As mentioned in the introduction, the very viability of a nonlinear diffusion is under scrutiny, for mechanical and rational reasons; but, analysis proves these functions are truly meaningless. 

\quotation{It illustrates
an empirical truth called "the BERS principle": GOD watches over
applied mathematicians. Let us hope he continues to do so \cite{Truesdell}.}

\section{Theorems}
\begin{theorem}[DERIVATIVE OF A CONVOLUTION INTEGRAL]\label{theorem1} The derivative of a convolution integral is equivalent to the derivative of either function involved in the convolution; said differently, one may selectively apply the derivative to either function of interest.
 
 $$\frac{\partial}{\partial x}\left(g\ast f\right)(x)=f'\ast g=f\ast g',$$
 
\noindent where prime indicates the derivative with respect to the variable of integration.

\noindent See R.N. Bracewell \cite{Bracewell}.
\end{theorem}

\begin{theorem}[DERIVATIVE OF A CONVOLUTION INTEGRAL]\label{theorem2} The derivative of a convolution integral by any variable not involved in the integral distributes across each function involve in the convolution.

$$\frac{\partial}{\partial t}\left(f\ast g\right)(x)=f'\ast g+ f\ast g'$$

\noindent See R.N. Bracewell \cite{Bracewell}.
\end{theorem}

\begin{theorem}[CONVOLUTION THEOREM]\label{theorem3} The convolution of two functions results in the multiplication of each Fourier transform of the functions, \emph{viz.:}
\begin{equation*}
 f\ast g=FG
\end{equation*}
\noindent See R.N. Bracewell \cite{Bracewell}.
\end{theorem}

\bibliographystyle{unsrt}
\bibliography{sample}

\end{document}